Coma Cluster Ultra-Diffuse Galaxies Are Not Standard Radio Galaxies

Mitchell F. Struble
Department of Physics & Astronomy, University of Pennsylvania, 209 South 33rd Street, Philadelphia, PA 19104

ABSTRACT

Matching members in the Coma cluster catalogue of ultra-diffuse galaxies (UDGs, Yagi et al. 2016) from SUBARU imaging with a very deep radio continuum survey source catalogue of the cluster (Miller et al. 2009) using the Karl G. Jansky Very Large Array (VLA) within a rectangular region of ~ 1.19 degrees$^2$ centred on the cluster core reveals matches consistent with random. An overlapping set of 470 UDGs and 696 VLA radio sources in this rectangular area finds 33 matches within a separation of 25 arcsec; dividing the sample into bins with separations bounded by 5 arcsec, 10 arcsec, 20 arcsec and 25 arcsec finds 1, 4, 17 and 11 matches. An analytical model estimate, based on the Poisson probability distribution, of the number of randomly expected matches within these same separation bounds is 1.7, 4.9, 19.4 and 14.2, each respectively consistent with the 95 percent Poisson confidence intervals of the observed values. Dividing the data into five clustercentric annuli of 0.1$^o$, and into the four separation bins, finds the same result. This random match of UDGs with VLA sources implies that UDGs are not radio galaxies by the standard definition. Those VLA sources having integrated flux > 1 mJy at 1.4 GHz in Miller et al. (2009) without SDSS galaxy matches are consistent with the known surface density of background radio sources. We briefly explore the possibility that some unresolved VLA sources near UDGs could be young, compact, bright, supernova remnants of type Ia events, possibly in the intracluster volume.

1. INTRODUCTION

Ultra-diffuse galaxies (UDGs) in the Coma cluster (van Dokkum et al. 2015; Koda et al. 2015) are characterised as having large sizes, ~ 40 percent comparable to the Galaxy's, but low surface brightnesses with approximately exponential light profiles. They show no evidence of tidal distortions, which implies they have large masses, ~ $10^9$ - $10^{10}$ $M_\odot$, comprised of dark matter (DM) to survive within the cluster environment and some may be a magnitude larger (van Dokkum et al. 2015; Yagi et al. 2016; Amorisco, Monachesi, & White 2017). Measurements of weak lensing by 784 UDGs in 18 nearby clusters (Coma not included) constrains their average subhalo masses within 30 kpc to $\lesssim 10^{11}$ $M_\odot$, but cannot rule out masses comparable the Galaxy's (Sifón et al. 2017). A study of 54 Coma UDGs claim all have low globular cluster counts comparable to nearby dwarf galaxies, and so may be genuine dwarfs with relatively larger halo spins responsible for their extended sizes (Amorisco et al. 2017; Amorisco & Loeb 2016). However, van Dokkum et al. (2017) find higher globular cluster counts for two Coma UDGs having stellar velocity dispersions which are comparable to regular galaxies. For sixteen Coma UDGs observed with HST, and from the relation between total galaxy mass and globular cluster count (Harris et al. 2017), they estimate a median halo mass ~ 1.5 x $10^{11}$ $M_\odot$. van Dokkum et al. (2017) attribute differences between their globular clusters counts and Amorisco et al.'s (2017) to their selection techniques for globulars, and inclusion of galaxies with smaller effective radii ($r_e$ < 1.5 kpc) in their sample. While UDGs have a mass spectrum, initial estimates of their average total mass are somewhat inconsistent, but van Dokkum et al.'s (2017) value appears to be the current best median estimate. Koda et al. (2015) estimate Coma contains ~ $10^3$ UDGs which, with this median mass, implies they are a non-negligible mass component of the cluster.

UDGs' radial clustercentric distribution follows that of bright cluster galaxies, and the major axes of those within 0.625° (~ 1 Mpc) of the cluster's center are preferentially oriented toward it (Yagi et al. 2016). van Dokkum et al. (2015) claimed UDGs appeared to avoid the inner cluster region where intracluster light is detected, but Koda et al. (2015) find UDGs in the cluster's central region arising from their ~ 1 magnitude deeper surface brightness data, a fact relevant to their survival in cluster cores.

Rong et al. (2017) studied the formation of UDG-like galaxies in Millennium-II and Phoenix ΛCDM simulations finding that they can form in cluster outskirts at late times (compared to typical dwarfs) in host haloes ~ $10^{10}$ $M_\odot$, which consequently fall into clusters at later times. The late formation time coupled with high halo spin can explain their lack of tidal features. However, the axial ratio distribution of Coma UDGs is more consistent with a distribution of prolate spheroids than oblate spheroids, and so are more likely to be dispersion supported rather than rotationally supported, which correspond to lower-mass stellar systems formed in ΛCDM simulations (Burkert 2017). Rong et al. (2017) do not discuss either their likely prolate shapes or clustercentric distribution of position angles in their ΛCDM simulation study.

About 52 percent of the UDGs are nucleated based on better two-component fits compared to single component fits of their light profiles. They lie along the red sequence of the colour-magnitude diagram of brighter Coma members, and show no evidence of H α emission, implying that their stellar component is a passively evolving population whose gas has been removed (Yagi et al. 2016). A recent spectroscopic study of five UDGs in the Coma field (Kadowaki et al. 2017) finds all are comprised of metal-poor stellar populations, but one has [OII] and [OIII] emission, a signature of star formation (Moustakas et al. 2006); one UDG was found to be a background galaxy.

Radio emission from some UDGs is found in low density, non-cluster environments, such as H I detected in isolated UDGs (Papastergis, Adams, & Romanowsky 2017). One out of four spectroscopically investigated UDGs that are Coma members do show evidence of star formation (Kadowaki et al. 2017), and thus may also show radio emission from HII or HI regions. Whether Coma cluster UDGs are radio sources due to AGN activity and/or DM annihilation (Colafrancesco et al. 2006, 2007) is also an interesting question. These questions motivate a preliminary test of such hypotheses by investigating matches of UDGs with radio sources.

2. DATA CATALOGUES USED

We use the very deep 1.4 GHz VLA radio continuum survey source catalogue of the cluster of Miller, Hornschemeier and Mobasher (2009, MHM), who observed two fields, Coma 1 corresponding to the cluster core, and Coma 3 corresponding to the southwest infall region. This survey confidently detected radio sources reaching 0.022 mJy (i.e., the minimum rms peak flux density), or a luminosity of $1.3 \times 10^{30}$ W Hz$^{-1}$ at an assumed cluster distance of 100 Mpc. MHMs' catalogue contains 1202 sources, of which 62.1 percent are unresolved. Each rectangular survey field covered ~ $1.06^\circ$ x $1.13^\circ$ with some overlap; these angular sizes were determined from the extrema of RA and DEC of catalogue listings. MHM list 628 sources in Coma 1, but the overlap region contains 68 sources listed in Coma 3, so the total number of sources within the RA and DEC bounds defined as Coma 1 is 696, covering an area of 1.195 degrees$^2$.

The UDG catalogue of Yagi et al. (2016), obtained from archival Subaru Prime Focus Camera imaging, covers a larger area, ~ $1.3^\circ$ x $2.7^\circ$, and completely overlaps the area of Coma 1 of the VLA radio catalogue and part of the area of Coma 3; it contains 854 UDGs. There are 470 UDGs within the VLA Coma 1 area defined above.

3. SEPARATION CRITERION FOR CATALOGUE MATCHES

MHM found matches with 499 bright Coma cluster galaxies in the SDSS DR5 catalogue (Adelman-McCarthy et al. 2007) using a separation criterion of 2 arcsec between optical and radio centroids for sources with *r* magnitude $\leq$ 22, and 1 arcsec for fainter sources. There are exceptions because of the inclusion of extended radio sources: their maximum listed offset in the catalogue between radio and optical SDSS position is 2.53 arcsec; there are 8 sources without a listed offset (of which 7 are extended sources) that are offset by up to ~ 15 arcsec from the optical galaxy as computed from the given coordinates. These larger offsets are caused by the centroid of an extended radio lobe, or lobes, as some are typical double-lobed radio galaxies, or are due to asymmetries in gas distribution expected for ram pressure stripping (see MHM's Fig. 5).

We considered a separation criterion of 15 arcsec from the above-mentioned maximum offset between radio and SDSS positions in MHM, but decided to use the approximate maximum FWHM of 25 arcsec of Subaru-UDGs in the COMA22 field of Yagi et al.'s (2016) catalog as a separation criterion between positions in the UDG and VLA catalogues to form a list of matches.

## 4. RESULTS & CONCLUSIONS

Catalogue comparisons yielded 1 match within 5 arcsec, 4 within 5 - 10 arcsec, 17 within 10 - 20 arcsec, and 11 within 20 - 25 arcsec. Twelve of the 33 UDGs are nucleated. Among the unique VLA source matches 23 are unresolved, and one VLA source, C1A-102, matches two UDGs. Table 1 lists the five UDG's within 10 arcsec of a VLA catalogue source, all in Coma 1, their separations $s$ in arcsec, UDG effective radius $r_e$ in kpc, the ratio of $s$ in kpc using Yagi et al.'s (2016) distance to Coma of 97.65 Mpc (2.114 arcsec = 1 kpc) to $r_e$, 20 cm peak flux density $F_{peak}$ in mJy, integrated 20 cm flux density $F_{int}$ in mJy, and relevant comments. The peak flux density at 1.4 GHz is nearly the same as the integrated flux density for all five sources in Table 1, which implies these sources are unresolved at MHM's survey resolution. VLA-UDG matches between 10 and 25 arcsec are listed in Table 2; $F_{int}$ ranges from 0.135 to 3.332 mJy. None of our matches lie within the 2 arcsec criterion adopted by MHM for the separation of SDSS galaxies and VLA sources; 11 of the VLA sources were matched with SDSS galaxies in MHM, all with $s \leq 1.65$ arcsec. There were only three UDG matches between 10 and 25 arcsec for the Coma 3 field; one is within the Coma 1 area by the RA and DEC bounds defined in Section 2, and the remaining two are ~ 1$^\circ$ from the cluster centre. None of the UDGs in Table 1 or 2 are in the small spectroscopic sample of Kadowaki et al. (2017).

**TABLE 1.** UDG and VLA source separations $\leq$ 10 arcsec in Coma 1 field (MHM)

| UDG | VLA source | $s$ arcsec | $r_e$ kpc | $s/r_e$ | $F_{peak}$ mJy | $F_{int}$ mJy | Comments; unresolved VLA sources not indicated |
|---|---|---|---|---|---|---|---|
| 181 | C1A-154 | 6.98 | - | - | 0.296 | 0.307 | no UDG Sersic parameters determined; MHM matches SDSS galaxy, $s$ = 0.57 arcsec |
| 272 | C1C-114 | 3.17 | 1.40 | 1.05 | 0.410 | 0.415 | |
| 308 | C1B-057 | 5.07 | - | - | 0.189 | 0.215 | *; no UDG Sersic parameters determined |
| 382 | C1B-141 | 8.49 | 1.48 | 2.71 | 0.232 | 0.291 | UDG nucleated; MHM matches SDSS galaxy, $s$ = 0.58 arcsec |
| 425 | C1B-171 | 9.07 | 3.11 | 1.38 | 0.121 | 0.133 | UGC also matched in Table 2; globular cluster counts in van Dokkum et al. (2017) |

\* Resolved source (MHM)

**TABLE 2.** UDG and VLA source separations between 10 and 25 arcsec in Coma 1 field (MHM)

| UDG | VLA Source | $s$ arcsec | Comments; unresolved VLA sources not indicated |
|---|---|---|---|
| 45  | C1C-190 | 22.90 | UDG nucleated; MHM matches SDSS galaxy, $s$ = 0.80 arcsec |
| 48  | C1C-929 | 21.49 | UDG nucleated |
| 86  | C1B-902 | 22.79 | *, no UDG Sersic parameters determined |
| 126 | C1B-229 | 18.62 | |
| 134 | C1B-237 | 16.85 | |
| 142 | C1B-247 | 18.97 | * |
| 245 | C1C-068 | 21.35 | *, MHM matches SDSS galaxy, $s$ = 0.15 arcsec |
| 248 | C1C-933 | 19.80 | UDG nucleated |
| 254 | C1C-084 | 18.97 | *, UDG nucleated |
| 287 | C1C-141 | 18.86 | |
| 303 | C1B-049 | 16.16 | |
| 316 | C1B-068 | 22.72 | UDG nucleated; MHM matches SDSS galaxy, $s$ = 1.07 arcsec |
| 361 | C1B-109 | 14.44 | UDG nucleated; MHM matches SDSS galaxy, $s$ = 0.12 arcsec |
| 371 | C1B-118 | 15.62 | UDG nucleated |
| 374 | C1B-129 | 16.99 | no UDG Sersic parameters determined |
| 405 | C1B-155 | 24.12 | UDG nucleated |
| 417 | C1B-167 | 13.21 | |
| 420 | C1A-929 | 22.03 | UDG nucleated; MHM matches SDSS galaxy, $s$ = 0.54 arcsec |
| 425 | C1B-933 | 15.66 | UDG also matched in Table 1; globular cluster counts in van Dokkum et al. (2017) |
| 426 | C1C-157 | 17.60 | no UDG Sersic parameters determined; MHM matches SDSS galaxy, $s$ = 0.41 arcsec |
| 452 | C1A-065 | 22.79 | |
| 460 | C1A-074 | 19.33 | MHM matches SDSS galaxy, $s$ = 1.30 arcsec |
| 468 | C1A-102 | 20.34 | MHM matches SDSS galaxy, $s$ = 1.65 arcsec |
| 469 | C1A-102 | 24.26 | same VLA source as UDG 468, and same SDSS galaxy match, $s$ = 1.65 arcsec |
| 639 | C1B-011 | 12.67 | *, UDG nucleated |
| 673 | C1B-056 | 15.95 | * |
| 674 | C1B-050 | 22.68 | * |
| 247 | C3A-146 | 18.25 | *, UDG nucleated; w/in Coma 1 area as defined in section 2 of text |
| 217 | C3C-132 | 11.11 | *, UDG nucleated |
| 514 | C3C-126 | 21.82 | no UDG Sersic parameters determined |

*Resolved source (MHM)

The smallest VLA-UDG separation in Table 1 (UDG 272) is ~ $r_e$, and the VLA source is the brightest of those within 10 arcsec separation. The source is unresolved and so is unlikely to arise from extended H I or synchrotron emission from electrons in a galactic magnetic field. We briefly explore the possibility that it might be a bright point source in, or near, the UDG, like a young supernova remnant (SNR).

Data on intracluster SN in groups and clusters have been identified as Type I or Ia. The first detection of a type I SN on an intergalactic bridge of a galaxy pair in a group (Rudnicki & Zwicky 1967), and the more recent detections of hostless intracluster SN Ia in Abell clusters (Gal-Yam et al. 2003; Graham et al. 2015), clearly indicate SNRs should be present in intracluster volumes arising from SN Ia progenitors lost from their parent galaxies, including the old stellar population of UDGs. Gal-Yam et al. (2003) estimate that ~ 20 percent of SN Ia are hostless within intracluster volumes. SNRs of SN Ia could possibly could be detected in Coma if they are bright enough, i.e. very young (only decades old), which is unlikely as optical surveys would have detected the outbursts. Early searches of SNR radio emission from known extragalactic SN Ia have yielded only upper limits (Sramek & Weiler 1990), and a more recent VLA survey of radio emission from known SN Ia, including one in a group in the cluster's far outskirts (iPTF 14atg in IC 831, distance ~ 90 Mpc), has not changed this conclusion (Chomiuk et al. 2016).

Galactic SNRs of SN Ia have diameters ~ 5 - 10 pc (Cox 2000), so at Coma's distance would only be ~ 1 - 2 x $10^{-2}$ arcsec in diameter, which would be unresolved in HMH's catalogue even if detectable. An estimate of the integrated flux density of an old SNR in Coma may be found from data of Kepler's SN Ia remnant: it has an integrated flux density of 19 Jy at 1 GHz, a spectral index of 0.64 (Green 2014), and distance 4.8 - 6.4 kpc (Reynoso and Goss, 1999). Scaling its integrated flux density to Coma's distance yields 3.7 - 6.6 x $10^{-5}$ mJy, which is far below MHM's detection limit. SNR G1.9+0.3, believed to be a SN Ia outburst ~ 150 years old near the galactic centre, has an integrated flux density of 935 mJy at 1.425 GHz in 2008 (it has brightened since discovery), and a spectral index of 0.62 (Green et al. 2008). Scaling its integrated flux density to Coma's distance yields ~ 7 x $10^{-6}$ mJy, again below MHM's detection limit. Chomiuk et al.'s (2016) searches for radio emission from SN Ia within 1 year of outburst lists results, both new and archival, for several at 1.4 GHz. Using their distances to host galaxies of this subsample finds estimated integrated flux density upper limits at Coma ranging from 0.003 mJy (SN 1985A, 46-75 days post-outburst) to 0.14 mJy (SN 2005gj, 166 days post-outburst). A SNR having the latter actual flux density could be detected in MHM's survey as it is above the minimum integrated flux density of 0.022 mJy for sources in their catalogue. Thus, any source bright enough to be detected in these radio surveys is also young enough that an optical counterpart would likely have been detected by supernova surveys prior to and during the time span of MHM's survey (late June 2006).

Inspection of the online Open Supernova Catalog (Guillochon et al. 2017) reveal one (SN2006bz, type Ia,) detected in Coma's RA and DEC ranges of both MHMs survey and Yagi et al.'s catalog, and within the cluster's redshift range, between one year prior to MHM's survey time, June 2005, and one month after their survey time, July 2006, to account for an initial premaximum increase of a SN outburst; it is associated with bright member galaxy. There are two background galaxies hosting SN in the Coma field: SN2005ck, type Ia, 9.5 arcsec (~ 12.8 kpc) from the nearest bright background galaxy (z = 0.08) within the RA and DEC ranges of both MHMs survey and Yagi et al.'s catalog; SN2006cj, type Ia, in a bright galaxy (z = 0.068) within RA and DEC range of MHMs survey but not of Yagi et al.'s catalog. No UDGs or radio sources are near any of these SN positions; none of the SN are in Chomiuk et al.'s (2016) searches. We conclude that none of the VLA radio sources in Tables 1 and 2 are likely associated with a SNR.

Comparing resolved vs. unresolved VLA sources in Tables 1 and 2 with their respective fractions in the total number of Coma 1 sources using a $\chi^2$-test of a 2X2 contingency table finds a $p$-value of 0.053, so the two samples are statistically indistinguishable at $p < 0.05$.

Comparing nucleated vs. unucleated UDGs in Tables 1 and 2 with their respective fractions in the total UDG sample using a $\chi^2$-test of a 2X2 contingency table finds a *p*-value of 0.073, which is not significant at $p < 0.05$. The number of nucleated and unucleated UDGs listed in Tables 1 and 2 are randomly selected from the total sample of UDGs.

Since none of our matches within 5 arcsec are large enough to be extended H I or synchrotron emission, and/or are associated with any UDG's nuclear region hosting AGN, they fail the criteria for the standard definitions for the physical origins of a radio galaxy based on morphology. We turn to the possibility that the entire set of matches within $s \leq 25$ arcsec are just random occurrences.

An estimate of the probability that the observed matches of UDGs with radio sources in a cluster are random may be obtained from application of the Poisson probability distribution. For a random distribution of galaxies with average surface density $n_{UDG}$, the probability that the single nearest radio source lies within the angular distance 0 to *s* is approximately (Rood 1976, 1979):

$$G(s) = 1 - \exp(-n_{UDG} \pi s^2). \quad (1)$$

The number of random matches of VLA sources, $N_{random}$, in a sample of $N_{VLA}$ sources expected to lie within *s* is then:

$$N_{random} = N_{VLA} \, G(s). \quad (2)$$

This formulation implicitly assumes a uniform distribution of both UDGs and VLA sources within a given surface area. We first assume uniform distributions of both samples within Coma 1.

There are 470 UDGs in Coma 1, so $n_{UDG} = 3.035 \times 10^{-5}$ arcsec$^{-2}$. The expected number of matches of VLA sources in the sample of $N_{VLA} = 696$ in the same area as the UDGs is ~ 1.7 for $s \leq 5$ arcsec, ~ 4.9 for $5 < s \leq 10$ arcsec, ~ 19.4 for $10 < s \leq 20$ arcsec, and ~ 14.2 for $20 < s \leq 25$ arcsec. Comparing these predicted values with our observed values (1, 4, 17, 11) within the same separation bounds indicates each are consistent with the 95 percent Poisson confidence intervals of the observed values, suggesting that all of our matches are random, supporting the conclusion that none of the UDGs in the Coma 1 field, as defined above, host radio sources. Since UDGs do not show any AGN-like radio emission and/or extended radio emission, which are criteria defining standard radio galaxies with implied physical origins, we amplify our conclusion to state that UDGs are not standard radio galaxies. Three UDGs are matched with *two* VLA sources for $s \leq 25$ arcsec; an estimate of paired VLA source matches expected randomly is $N_{random\ pairs} \approx N_{VLA} [G(s)]^2$, or 2.3, consistent within Poisson uncertainty.

UDGs decrease in surface density from the cluster centre, so we estimate the number of random sources expected as a function of clustercentric radius. We bin the samples into five clustercentric annuli of 0.1° width, assuming Yagi et al's (2014) centroid, and then apply Equations (1) and (2) to the respective subsamples of binned separations. There are 32 matches within the area (the one Coma 3 VLA source within the Coma 1 area is > 0.5° from the cluster centre); Table 3 presents the results. Comparing predicted with observed values in each annulus, and within the same separation bounds, indicates each are consistent with the 95 percent Poisson confidence intervals of the observed values. For completeness, the final row of Table 3 assumes a uniform surface density over the circular area of 0.5° radius; the results are consistent with random expectation, within 95 percent Poisson confidence intervals of the observed values.

**TABLE 3.** Clustercentric Number Counts, and Observed and Predicted Random Values, within Separation Bins via Equations (1) and (2)

| Annulus degrees | $N_{UDG}$ | $N_{VLA}$ | $N_{obs}$ in $s$ bin arcsec interval | | | | $N_{random}$ in $s$ bin arcsec interval | | | |
|---|---|---|---|---|---|---|---|---|---|---|
| | | | 0-5 | 5-10 | 10-20 | 20-25 | 0-5 | 5-10 | 10-20 | 20-25 |
| 0 - 0.1 | 14 | 22 | 0 | 0 | 1 | 0 | 0.06 | 0.18 | 0.73 | 0.53 |
| 0.1 - 0.2 | 81 | 89 | 0 | 2 | 4 | 1 | 0.45 | 1.38 | 5.28 | 3.75 |
| 0.2 - 0.3 | 99 | 146 | 1 | 0 | 6 | 4 | 0.56 | 1.64 | 6.44 | 4.65 |
| 0.3 - 0.4 | 90 | 147[a] | 0 | 2 | 4 | 5 | 0.36 | 1.09 | 4.25 | 3.11 |
| 0.4 - 0.5 | 99 | 135[b] | 0 | 0 | 1 | 1 | 0.28 | 0.97 | 3.15 | 2.47 |
| 0 - 0.5 | 383 | 539[a,b] | 1 | 4 | 16 | 11 | 1.6 | 4.73 | 18.54 | 13.48 |

[a] Includes 3 sources in Coma 3
[b] Includes 14 sources in Coma 3

About half of all radio sources in the Coma 1 field of MHM are unassociated with SDSS galaxies that are cluster members, and so are likely background radio sources. The surface density of radio sources is ~ 89.5 degree$^{-2}$ for sources > 1 mJy at 1.4 GHz (Helfand, White & Becker 2015), so in the area of Coma 1 defined in Sec. 2 one expects to find ~ 107 background radio sources of integrated flux > 1 mJy at 1.4 GHz. MHM's catalogue contains 172 sources of integrated flux > 1 mJy in Coma 1, of which 94 are not matched with an SDSS galaxy, which is consistent with the 95 percent Poisson confidence intervals of the known surface density of background radio sources.

There may, however, be an exception to the above conclusions. One UDG in van Dokkum et al.'s (2017) study, No. 425 in Yagi et al's (2016) catalog, is in both Tables 1 and 2 as it is matched with two unresolved radio sources of approximately similar $F_{int}$ (in Table 2: C1B-933, $F_{int}$ = 0.153 mJy) at angular separations of 9.07 and 15.66 arcsec (~ 4.3 and 7.4 kpc, respectively), and are separated by 24.62 arcsec from each other. Positionally these sources are approximately antipodal with respect to, but not equidistant from, UDG 425. This is suggestive of non-thermal radio galaxy morphology: a pair of faint unresolved radio lobes, or putative hotspots within undetected fainter lobes, astride an optical galaxy. The UDG has a single Sersic photometric fit with $r_e$ = 3.11 arcsec (~ 1.5 kpc), and is one of the bluer UDGs ($B - R$ = 0.6) in the cluster core. It has an average globular cluster count (24 $\pm$ 11) in van Dokkum et al.'s (2017) study, implying a total mass ~ 1.5 x 10$^{11}$ $M_\odot$, their median value. Possible radio properties of UDG 425 must await observations deeper than MHM's survey. If future investigations find this UDG to be a *bona fide*, but faint, radio galaxy, it does not change the conclusion that the great majority of Coma's UDGs are not radio galaxies by the standard definition.

Searches for radio emission from DM-dominated local dwarf spheroidal galaxies (dSphs) by Spekkens et al. (2013), which could be considered small, low-mass versions of UDGs since they have analogous physical characteristics, including likely prolate DM halos (Hayashi & Chiba 2015), find none. Models of radio emission via these processes by Colafrancesco et al. (2015) predict flux densities ~ 1 mJy at 1.4 GHz for galaxies ~ 10$^{12}$ $M_\odot$ at redshift ~ 0.01. One Coma cluster UDG, Dragonfly 44, has an estimated mass ~ 10$^{12}$ $M_\odot$ (van Dokkum et al. 2016), and is No. 11 in Yagi et al.'s (2016) UDG catalogue, but is outside MHM's survey areas. Estimates of total median mass of Coma UDGs of 1.5 x 10$^{11}$ $M_\odot$ from globular cluster counts (van Dokkum et al. 2017) would predict flux densities ~ 0.03 mJy at 1.4 GHz from DM annihilation at Coma's distance (Colafrancesco et al.

2015, Fig. 3), which might be marginally detected in MHM's survey. We thus cannot draw confident conclusions regarding the absence of radio synchrotron emission from Coma's UDGs via DM annihilation products within extended magnetic fields as did Spekkens et al. (2013) for dSphs.

ACKNOWLEDGEMENTS

Part of this work is based on archival data, software or online services provided by the ASI SCIENCE DATA CENTER (ASDC). The 95 percent Poisson confidence intervals were computed from exact expressions for small counts used at http://statpages.info/confint.html. I thank James Aguirre for technical discussions and clarifications, Saul Kohn for technical discussions, and Kenneth Cooper for software assistance. I also thank the reviewer for very helpful comments, particularly for directing attention to a recent study indicating upward revision of UDG median total mass estimate.